\documentstyle[twoside,psfig,graphicx]{article}

\catcode`\@=11
\long\def\@makefntext#1{
\protect\noindent \hbox to 3.2pt {\hskip-.9pt
$^{{\eightrm\@thefnmark}}$\hfil}#1\hfill}       

\def\@makefnmark{\hbox to 0pt{$^{\@thefnmark}$\hss}}    

\def\ps@myheadings{\let\@mkboth\@gobbletwo
\def\@oddhead{\hbox{}
\rightmark\hfil\eightrm\thepage}
\def\@oddfoot{}\def\@evenhead{\eightrm\thepage\hfil
\leftmark\hbox{}}\def\@evenfoot{}
\def\sectionmark##1{}\def\subsectionmark##1{}}

\oddsidemargin=\evensidemargin
\newcounter{sectionc}\newcounter{subsectionc}\newcounter{subsubsectionc}
\renewcommand{\section}[1] {\vspace{12pt}\addtocounter{sectionc}{1}
\setcounter{subsectionc}{0}\setcounter{subsubsectionc}{0}\noindent
    {\tenbf\thesectionc. #1}\par\vspace{5pt}}
\renewcommand{\subsection}[1] {\vspace{12pt}\addtocounter{subsectionc}{1}
    \setcounter{subsubsectionc}{0}\noindent
    {\bf\thesectionc.\thesubsectionc. {\kern1pt \bfit #1}}\par\vspace{5pt}}
\renewcommand{\subsubsection}[1] {\vspace{12pt}\addtocounter{subsubsectionc}{1}
    \noindent{\tenrm\thesectionc.\thesubsectionc.\thesubsubsectionc.
    {\kern1pt \tenit #1}}\par\vspace{5pt}}
\newcommand{\nonumsection}[1] {\vspace{12pt}\noindent{\tenbf #1}
    \par\vspace{5pt}}

\newcounter{appendixc}
\newcounter{subappendixc}[appendixc]
\newcounter{subsubappendixc}[subappendixc]
\renewcommand{\thesubappendixc}{\Alph{appendixc}.\arabic{subappendixc}}
\renewcommand{\thesubsubappendixc}
    {\Alph{appendixc}.\arabic{subappendixc}.\arabic{subsubappendixc}}

\renewcommand{\appendix}[1] {\vspace{12pt}
        \refstepcounter{appendixc}
        \setcounter{figure}{0}
        \setcounter{table}{0}
        \setcounter{lemma}{0}
        \setcounter{theorem}{0}
        \setcounter{corollary}{0}
        \setcounter{definition}{0}
        \setcounter{equation}{0}
        \renewcommand{\thefigure}{\Alph{appendixc}.\arabic{figure}}
        \renewcommand{\thetable}{\Alph{appendixc}.\arabic{table}}
        \renewcommand{\theappendixc}{\Alph{appendixc}}
        \renewcommand{\thelemma}{\Alph{appendixc}.\arabic{lemma}}
        \renewcommand{\thetheorem}{\Alph{appendixc}.\arabic{theorem}}
        \renewcommand{\thedefinition}{\Alph{appendixc}.\arabic{definition}}
        \renewcommand{\thecorollary}{\Alph{appendixc}.\arabic{corollary}}
        \renewcommand{\theequation}{\Alph{appendixc}.\arabic{equation}}
        \noindent{\tenbf Appendix \theappendixc #1}\par\vspace{5pt}}
\newcommand{\subappendix}[1] {\vspace{12pt}
        \refstepcounter{subappendixc}
        \noindent{\bf Appendix \thesubappendixc. {\kern1pt \bfit #1}}
    \par\vspace{5pt}}
\newcommand{\subsubappendix}[1] {\vspace{12pt}
        \refstepcounter{subsubappendixc}
        \noindent{\rm Appendix \thesubsubappendixc. {\kern1pt \tenit #1}}
    \par\vspace{5pt}}

\newcommand{\textlineskip}{\baselineskip=13pt}
\newcommand{\smalllineskip}{\baselineskip=10pt}

\def\eightcirc{
\begin{picture}(0,0)
\put(4.4,1.8){\circle{6.5}}
\end{picture}}
\def\eightcopyright{\eightcirc\kern2.7pt\hbox{\eightrm c}}

\newcommand{\copyrightheading}[1]
    {\vspace*{-2.5cm}
    {
     }}

\def\abstracts#1#2#3{{
    \centering{\begin{minipage}{4.5in}\footnotesize\baselineskip=10pt
    \parindent=0pt #1\par
    \parindent=15pt #2\par
    \parindent=15pt #3
    \end{minipage}}\par}}

\newcommand{\bibit}{\nineit}

\renewenvironment{thebibliography}[1]
    {\frenchspacing
     \ninerm\baselineskip=11pt
     \begin{list}{\arabic{enumi}.}
    {\usecounter{enumi}\setlength{\parsep}{0pt}
     \setlength{\leftmargin 12.7pt}{\rightmargin 0pt} 
     \setlength{\itemsep}{0pt} \settowidth
    {\labelwidth}{#1.}\sloppy}}{\end{list}}

\newcounter{itemlistc}
\newcounter{romanlistc}
\newcounter{alphlistc}
\newcounter{arabiclistc}

\newcommand{\fcaption}[1]{
        \refstepcounter{figure}
        \setbox\@tempboxa = \hbox{\footnotesize Fig.~\thefigure. #1}
        \ifdim \wd\@tempboxa > 5in
           {\begin{center}
        \parbox{5in}{\footnotesize\smalllineskip Fig.~\thefigure. #1}
            \end{center}}
        \else
             {\begin{center}
             {\footnotesize Fig.~\thefigure. #1}
              \end{center}}
        \fi}

\newcommand{\tcaption}[1]{
        \refstepcounter{table}
        \setbox\@tempboxa = \hbox{\footnotesize Table~\thetable. #1}
        \ifdim \wd\@tempboxa > 5in
           {\begin{center}
        \parbox{5in}{\footnotesize\smalllineskip Table~\thetable. #1}
            \end{center}}
        \else
             {\begin{center}
             {\footnotesize Table~\thetable. #1}
              \end{center}}
        \fi}

\def\@citex[#1]#2{\if@filesw\immediate\write\@auxout
    {\string\citation{#2}}\fi
\def\@citea{}\@cite{\@for\@citeb:=#2\do
    {\@citea\def\@citea{,}\@ifundefined
    {b@\@citeb}{{\bf ?}\@warning
    {Citation `\@citeb' on page \thepage \space undefined}}
    {\csname b@\@citeb\endcsname}}}{#1}}

\newif\if@cghi
\def\cite{\@cghitrue\@ifnextchar [{\@tempswatrue
    \@citex}{\@tempswafalse\@citex[]}}
\def\citelow{\@cghifalse\@ifnextchar [{\@tempswatrue
    \@citex}{\@tempswafalse\@citex[]}}
\def\@cite#1#2{{$\null^{#1}$\if@tempswa\typeout
    {IJCGA warning: optional citation argument
    ignored: `#2'} \fi}}

\def\pmb#1{\setbox0=\hbox{#1}
    \kern-.025em\copy0\kern-\wd0
    \kern.05em\copy0\kern-\wd0
    \kern-.025em\raise.0433em\box0}

\def\fnt#1#2{\footnotetext{\kern-.3em
    {$^{\mbox{\scriptsize #1}}$}{#2}}}

\def\@makefnmark{\hbox to 0pt{$^{\@thefnmark}$\hss}}    

\def\ps@myheadings{%
    \let\@oddfoot\@empty\let\@evenfoot\@empty
    \def\@evenhead{\slshape\leftmark\hfil}
    \def\@oddhead{\hfil{\slshape\rightmark}}
    \let\@mkboth\@gobbletwo
    \let\sectionmark\@gobble
    \let\subsectionmark\@gobble
    }
\font\tenrm=cmr10
\font\tenit=cmti10
\font\tenbf=cmbx10
\font\bfit=cmbxti10 at 10pt
\font\ninerm=cmr9
\font\nineit=cmti9

\font\eightrm=cmr8

\def\qed{\hbox{${\vcenter{\vbox{            
   \hrule height 0.4pt\hbox{\vrule width 0.4pt height 6pt
   \kern5pt\vrule width 0.4pt}\hrule height 0.4pt}}}$}}

\begin{document}
\setlength{\textheight}{7.7truein}  

\thispagestyle{empty}

\markboth{\protect{\footnotesize\it V.N.Marachevsky
}}{\protect{\footnotesize\it Casimir Energy
of a Dilute Dispersive Dielectric Ball}}

\normalsize\textlineskip

\setcounter{page}{1}

\copyrightheading{}     

\vspace*{0.88truein}

\centerline{\bf CASIMIR ENERGY OF A DILUTE DISPERSIVE DIELECTRIC
 BALL{\Large:} }
\vspace*{0.035truein}
\centerline{\bf REALISTIC MICROSCOPIC MODEL}
\vspace*{0.37truein}
\centerline{\footnotesize VALERY N. MARACHEVSKY
\footnote{e-mail: root@VM1485.spb.edu, maraval@mail.ru}}
\baselineskip=12pt
\centerline{\footnotesize\it
Department of Theoretical Physics, St.Petersburg State University}
\baselineskip=10pt
\centerline{\footnotesize\it
198504 St.Petersburg, Russia}


\vspace*{0.21truein}
\abstracts{
The Casimir energy of a dilute homogeneous nonmagnetic dielectric
ball at zero temperature is derived analytically within a {\it microscopic}
realistic model of dielectrics for an arbitrary physically possible
frequency dispersion of dielectric permittivity $\varepsilon(i\omega)$.
Divergences are absent in calculations, a minimum interatomic
distance $\lambda$ is a {\it physical} cut-off. Casimir surface force
is proved to be attractive. A physical definition of the Casimir energy
is discussed.}{}{}


\vspace*{1pt}\textlineskip  
\section{Microscopic Approach}    
\vspace*{-0.5pt}
\noindent
Consider a dielectric nonmagnetic  ball of the radius $a$ and permittivity
$\varepsilon$ at zero temperature, surrounded  by  a vacuum.  The ball is dilute, i.e. all
final expressions  are obtained under the assumption $\varepsilon - 1
\ll 1$ in the order $(\varepsilon(i\omega) - 1)^2$, the lowest order that
yields the  energy of interaction between atoms of the ball.

The Casimir energy of a  disjoint macroscopic system
(two dispersive dielectric parallel plates is
a classic example by Lifshitz) depends only on the
distance between macroscopic bodies
and dispersion of dielectrics\cite{Lifshitz}.
On the other hand, it was first argued in Refs.\cite{2,3}
that for a dilute connected dielectric the Casimir energy at zero
temperature is equal to the energy of dipole-dipole pairwise interactions
of all atoms constituting the dielectric
and thus should also depend on an average minimum distance between
atoms of a dielectric $\lambda$.
For a dilute dispersive dielectric ball with an arbitrary
frequency dependent dielectric permittivity
the Casimir energy at zero temperature was first derived in Ref.\cite{4}.

A dipole-dipole interaction  of two neutral atoms with atomic
polarizabilities $\alpha_1(i \omega)$ and $\alpha_2(i \omega)$
is described by a potential \cite{Dzialoshinskii}
\begin{equation}
U(r) = - \frac{1}{\pi r^2} \int_{0}^{+ \infty} \,
\omega^4 \alpha_1(i \omega) \, \alpha_2 (i \omega)\,  e^{-2\, \omega r }
\Bigl[1+ \frac{2}{\omega r} + \frac{5}{(\omega r)^2 }
+ \frac{6}{(\omega r)^3} + \frac{3}{(\omega r)^4} \Bigr] d\omega , \label{tt1}
\end{equation}
where $r$ is a distance between two atoms. The energy calculation for
the ball
is illustrated by Fig.$1$. Suppose that an atom with
an atomic polarizability $\alpha(i\omega)$ is located at the point
$B$. One has to integrate interaction of an atom at the point $B$
via a potential (\ref{tt1})  with the atoms separated by distances
greater than interatomic distances $\lambda$ from the point $B$,
integrate over all atom locations $B$ inside the ball and multiply
by a factor $1/2$ to calculate the energy.

\begin{figure}[htbp] 
\vspace*{13pt}
\centering
\hspace{1cm}
\includegraphics[height=4.5cm]{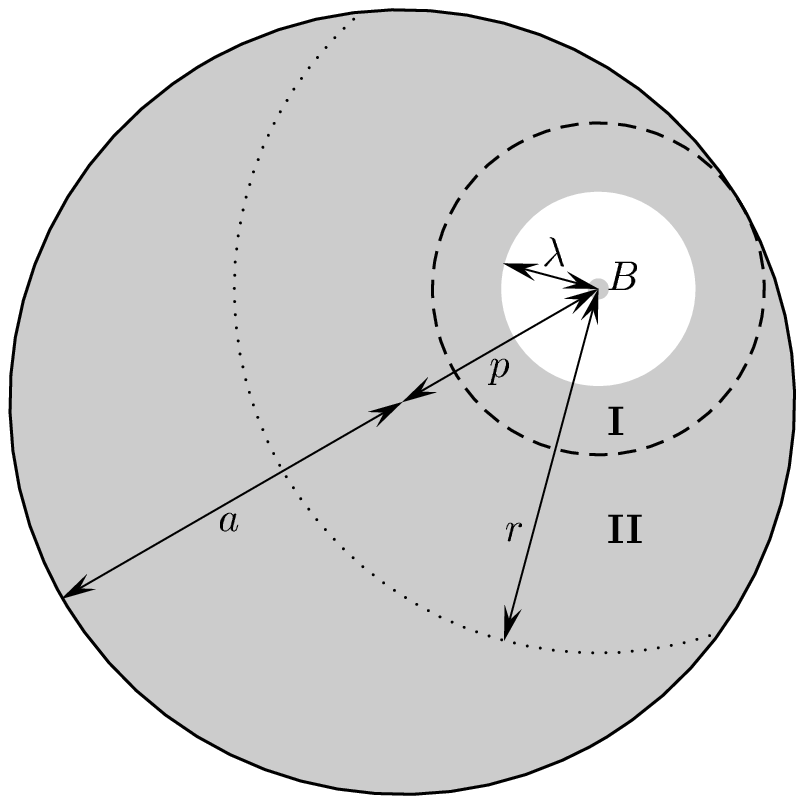} \hfill
\includegraphics[height=4.5cm]{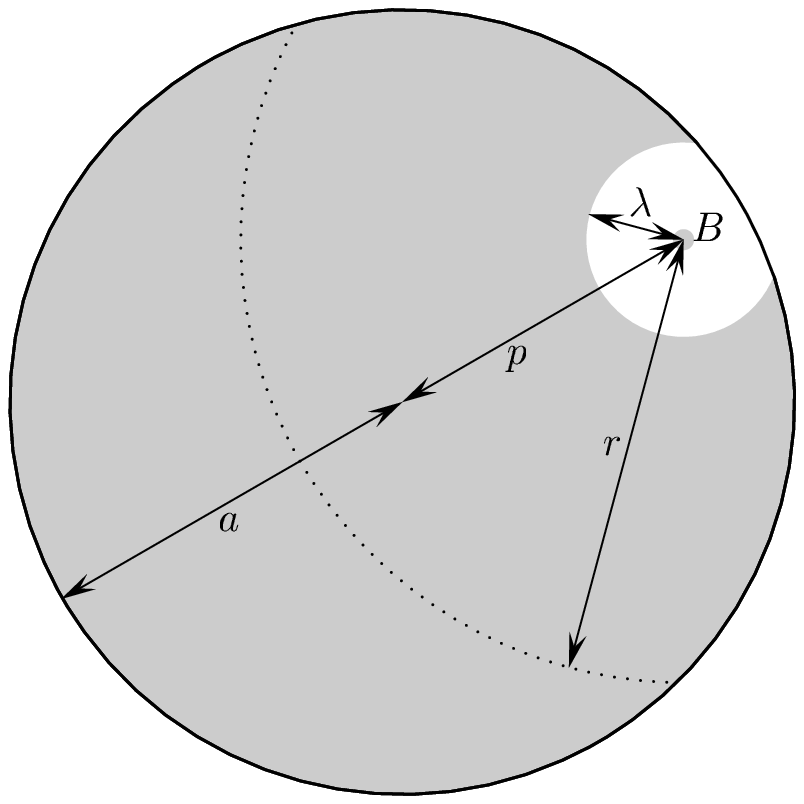} \hspace{1cm}
\vspace*{13pt}
\fcaption{ Energy calculation for the ball. Shaded
areas $\bf \Large I$ and $\bf \Large II$ separated by a dashed
line on the left picture denote areas of integration in the first
and second integrals in (\ref{tt2}) respectively. A shaded area on
the right picture represents the area of integration in the third
integral in (\ref{tt2}). A part of the sphere of radius $r$ which
is located inside the dielectric ball is denoted by a dotted line.
Its area is equal to $2\pi r^2 (1 - (p^2+r^2-a^2)/(2pr))$. }
\end{figure}

Assuming homogeneity of
the ball (it results in
$\alpha_1(i\omega)=\alpha_2(i\omega)=\alpha(i\omega)$ and the
condition that the number density of atoms $\rho$ doesn't depend
on the point inside the ball), the Casimir energy is equal to
\begin{eqnarray}
E &=& \frac{\rho^2}{2} \Biggl(\int_{0}^{a-\lambda} dp\, 4\pi p^2 \int_{\lambda}^{a-p}
dr\, 4\pi r^2 U(r) \nonumber \\& &+ \int_{0}^{a-\lambda} dp\, 4\pi p^2
\int_{a-p}^{a+p} dr \, 2\pi r^2 \Bigl(1-\frac{r}{2p}-
\frac{p^2-a^2}{2pr}\Bigr) U(r) \nonumber \\
& &+ \int_{a-\lambda}^{a} dp \, 4\pi p^2 \int_{\lambda}^{a+p} dr\,
2\pi r^2 \Bigl(1-\frac{r}{2p}- \frac{p^2-a^2}{2pr}\Bigr) U(r)
 \Biggr)  \label{tt2}
\end{eqnarray}

Performing calculations, it is possible to derive an analytic result\cite{4}:
\begin{eqnarray}
E&=& - \rho^2 \frac{\pi}{48} \int_{0}^{+\infty} d\omega \alpha^2(i\omega) \,
 \Bigl( \frac{a^3}{\lambda^3} e^{-2\omega\lambda} (128 + 256 \omega\lambda +
 128 \omega^2\lambda^2  + 64 \omega^3\lambda^3 ) - \nonumber \\
 & &-\frac{a^2}{\lambda^2} \bigl(e^{-2 \omega\lambda}(144 + 288\omega\lambda +
 120\omega^2\lambda^2 + 48\omega^3\lambda^3 ) - 96\omega^2\lambda^2
 E_1(2\omega\lambda) \bigr) + \nonumber \\& & \bigl(e^{-2\omega\lambda}(41 +
 34\omega\lambda + 14\omega^2\lambda^2 + 4\omega^3\lambda^3 ) +
24 E_1(2\omega\lambda)\bigr) + \nonumber \\& & \bigl(e^{-4\omega a} (-21 + 12\omega a) -
 E_1(4\omega a) (24 + 96 \omega^2 a^2) \bigr)   \Bigr) ,  \label{tt3}
\end{eqnarray}
where $E_1(x)=\int_{1}^{+\infty} e^{-tx}/t \, dt$.

Eq.(\ref{tt3}) finally solves the problem of the Casimir
energy for a dilute dielectric ball at zero temperature.
This energy is finite and physical only when a finite minimum separation
between atoms $\lambda$ is taken into account.

Some words need to be said about the change of a viewpoint on the
physical definition of the Casimir energy.
When an interatomic distance $\lambda$ is taken into
account, the Casimir energy can be calculated without
divergences by use of Eq.(\ref{tt2}), and so calculated energy doesn't require further
renormalization.  No further renormalization is needed since it is
obvious from the method of calculations that
the Casimir energy (\ref{tt3}) coincides with a potential (binding)
energy of the ball
when $\lambda$ is large enough, so that effects of short-range
interatomic repulsive forces and other interactions can be neglected.
In this case there is no classical part of the energy in the beginning
of calculations,
there exists only a set of atoms with
specified interactions between them due to quantum fluctuations.
The sum of all these
interactions between atoms should give the Casimir contribution to the
binding energy of a dielectric ball  -- a macroscopic classical system.

The leading contribution
from the last line in (\ref{tt3}) is equal to
\begin{eqnarray}
&-&\rho^2 \alpha^2(0) \frac{\pi}{48} \int_{0}^{+\infty} d\omega  \,
  \bigl(e^{-4\omega a} (-21 + 12\omega a)
-E_1(4\omega a) (24 + 96 \omega^2 a^2) \bigr)  \nonumber \\
 &=& \rho^2 \alpha^2(0) \frac{23}{96} \frac{\pi}{a} =
  \frac{23}{1536\pi a} (\varepsilon - 1)^2 = E_{ld}. \label{tt5}
\end{eqnarray}
The term (\ref{tt5}) can be called a contribution from large
distances to the  Casimir energy of the ball. Usually all terms
different from (\ref{tt5}) were simply discarded or added up to
macroscopic quantities of the ball (e.g., volume, surface
energies) during the renormalization procedure, as it often
happens in field theory. So only the term (\ref{tt5}) was usually
considered as the Casimir energy term. In fact, this was an
erroneous definition of the Casimir energy since no physical
quantities can be calculated by use of Eq.(\ref{tt5}). All
physical quantities that can be measurable at least in principle
for a dilute ball (e.g., surface force or trajectories of the
particles near the ball surface during the ball collapse in the
adiabatic approximation) have to be calculated by use of a total
potential energy of the ball (\ref{tt3}). The term  (\ref{tt5})
has little influence on physics. The reason is quite simple -  the
term (\ref{tt5}) is much less in magnitude than the terms in the
first, second and  third  lines of (\ref{tt3}). For example, ratio
of terms in the first line of (\ref{tt3}) to (\ref{tt5}) is $\sim
(\omega_0 a) (a/\lambda)^3 \gg 1$, where $\omega_0$ is a
characteristic atomic absorption frequency. However, the term
(\ref{tt5}) was really important for development of the theory of
Casimir effect in connected dielectrics since this term has been
derived via microscopic \cite{Milton} and macroscopic \cite{Mar1}
techniques - so the equivalence of large distance parts of the
Casimir energy for a dilute dielectric ball derived by microscopic
and macroscopic approaches was proved.

It is important to stress that so far macroscopic  methods did not yield
satisfactorily short distance contributions to the Casimir energy
of connected dielectrics\cite{Vas}. The reason is simple: these methods
were developed for {\it disjoint}, not connected dielectrics.

From (\ref{tt2}) it follows that Casimir surface force on a
dilute dielectric ball is attractive.
It is convenient to define
$N\equiv a/\lambda, p\equiv\omega\lambda$. Then Eq.(\ref{tt3})
can be rewritten in a general form
\begin{equation}
E = - \frac{\rho^2}{\lambda} \int_{0}^{+\infty} dp \, \alpha^2\Bigl(i
\frac{p}{\lambda}\Bigr) f(N,p) .   \label{tt6}
\end{equation}
The function $f(N,p)>0$ for $N> 1/2, p>0$. Conservation of atoms
inside the ball and homogeneity impose the condition
\begin{equation}
\rho \, \frac{4\pi a^3}{3} = const. \label{tt8}
\end{equation}
From the condition of an atomic conservation (\ref{tt8}) it
follows that during the ball collapse or expansion
\begin{equation}
N={\rm const}. \label{tt7}
\end{equation}
It is convenient to use  Kramers--Kronig relations in the form
\begin{equation}
\alpha(i\omega) = \int_{0}^{+\infty} dx\, \frac{x g(x)}{x^2+\omega^2}, \label{tt9}
\end{equation}
where the condition $g(x)>0$ always holds. Using (\ref{tt6}),
(\ref{tt8}), (\ref{tt7}), (\ref{tt9}), Casimir force on a unit
surface is equal to
\begin{eqnarray}
 F &=& -\frac{1}{4\pi a^2} \frac{\partial E}{\partial a} \nonumber \\ &=&
-\frac{\rho^2}{4\pi a^3}\int_{0}^{+\infty} d\omega \int_{0}^{+\infty}
dx  \frac{x (7x^2 + 3 \omega^2) g(x)}{(x^2+\omega^2)^2}\,
\alpha(i\omega) f(N,\omega \lambda)  < 0.  \label{tt10}
\end{eqnarray}
$F<0$ because all functions inside integrals are positive. Casimir
surface force is attractive for every model of atomic
polarizability consistent with general causal requirements.

\nonumsection{References}

\eject

\end{document}